\documentclass{nuws}

\def\backsect{\vspace*{-3mm}}

\usepackage{times}
\usepackage{graphicx} 
\begin{document}

\title{Accuracy of muon transport simulation}
\author[1,2]{Igor~A.~Sokalski} 
\affil[1]{DAPNIA/SPP, CEA/Saclay, 91191 Gif-sur-Yvette Cedex, France}
\author[2]{Edgar~V.~Bugaev}
\author[2]{Sergey~I.~Klimushin}
\affil[2]{Institute for Nuclear Research, Russian Academy of Science,
60th October Anniversary prospect 7a, Moscow 117312, Russia}

\correspondence{sokalski@hep.saclay.cea.fr,\\sokalski@pcbai10.inr.ruhep.ru}

\firstpage{1}
\pubyear{2001}

\maketitle

\begin{abstract}
Chain of calculations which have to be performed to predict any kind of signal 
in a deep underwater/ice neutrino detector necessarily includes the lepton 
propagation through thick layers of matter, as neutrino can be observed only 
by means of leptons (muons, first of all, due to their large ranges) that are 
generated in $\nu\,N \rightarrow l\,N$ interactions. Thus, the muon 
propagation plays a key role when analyzing data and it is important to 
understand clearly how transportation part of simulation chain contributes to 
total inaccuracy of final results. Here we consider sources of 
uncertainties that appear in Monte Carlo algorithms for simulation of muon 
transport. The trivial but effective test is proposed to measure the 
propagation algorithm accuracy. The test is applied to three MC muon transport 
codes (PROPMU, MUSIC, MUM) and results are reported.
\end{abstract}

\section{Introduction}

The main challenge for existing and planned underwater/ice neutrino 
telescopes \cite{amanda1},\cite{amanda2},\cite{antares1},\cite{antares2},
\cite{baikal1},\cite{baikal2},\cite{nestor1},\cite{nestor2} is detection
of neutrino of extraterrestrial origin. But neutrino is neutral weakly 
interacting particle and one can not observe it directly but has to
detect it by means of lepton that appears in $\nu\,N \rightarrow l\,N$ 
interaction and passes a distance in medium before being detected. The case
with $l\,=\mu^{\pm}$ is considered, first of all, because muons possess the 
best ability to propagate large  distances in 
contrast to $e^{\pm}$ (due to their short radiation length) and $\tau^{\pm}$ 
(due to their short life time) in a wide range of energies. On the other hand, 
atmospheric muons (that represent a penetrating component of atmospheric 
showers resulting from interaction of primary cosmic radiation with the Earth
atmosphere) are the principal background for neutrino signal. To eliminate
it one has to predict correctly the detector response on atmospheric muon
flux. Also one should keep in mind the fact that atmospheric muons are of 
scientific interest themselves: for instance, from the point of view
of charm production \cite{prompt1}. The last (but not the least) important
point is detector calibration which has to be done, again by means of 
atmospheric muons that are the only more or less known intensive 
calibration source for deep underwater/ice neutrino telescopes. 

Thus, when analyzing experimental data obtained with underwater/ice detectors 
one has to apply some model for muon transport through thick layers of 
matter, using an algorithm that gives the muon energy $E_{1}$ at the end of 
distance $D$ providing that its initial energy is equal to $E_{0}$. Commonly, 
Monte Carlo (MC) technique is used for this purpose \cite{fluka1}, 
\cite{lipari},\cite{music1},\cite{music2},\cite{mum1},\cite{mmc1} being the 
most adequate to essentially stochastic nature of muon energy losses. Such a 
MC algorithm necessarily incorporates some uncertainties into the final 
result. These uncertainties can be divided into two parts:

\begin{itemize}
\item[\bf{(A)}] insurmountable uncertainties which relates to
\begin{itemize}
\item  finite accuracy of formulae for muon cross-sections;
\item  data on medium density and composition.
\end{itemize}
\item[\bf{(B)}] ``inner'' errors that are produced by simulation algorithm 
               itself due to
\begin{itemize}
\item  finite accuracy of numerical procedures;
\item  simplifications that are done to get reasonable 
       computation time. 
\end{itemize}
\end{itemize}

\begin{figure}[h]
\vspace*{0.0mm} 
\includegraphics[width=8.0cm]
{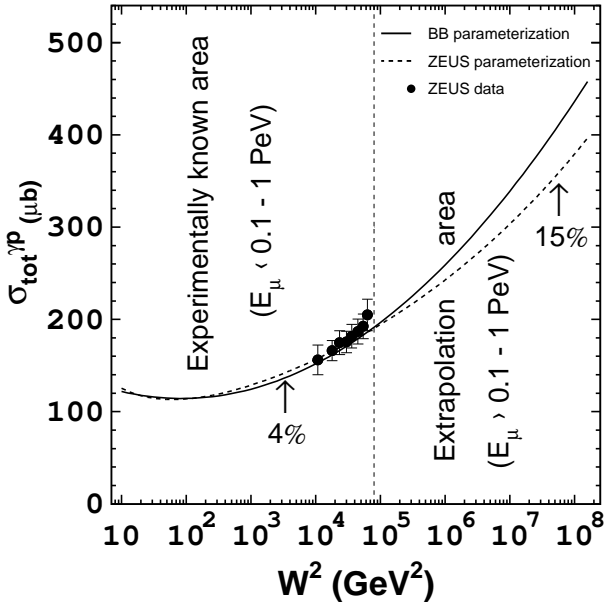} 
\label{fig:sigma_exp}
\caption{
Total cross-section $\sigma^{\gamma p}_{tot}$ as a function of $W^{2}$ as 
measured and parametrized by ZEUS Collaboration (markers and dashed 
line, correspondingly), and parametrized by Bezrukov-Bugaev (solid 
line).
}
\end{figure}
\begin{figure}[h]
\vspace*{-2.0mm} 
\includegraphics[width=8.0cm]
{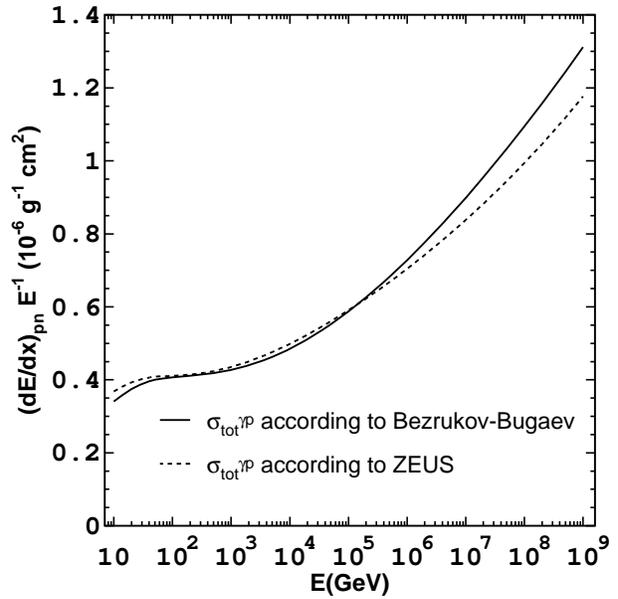} 
\label{fig:enlopn}
\caption{
Muon energy losses due to photonuclear interaction in pure water as 
obtained using (a) BB parameterization for  $\sigma^{\gamma p}_{tot}$ 
(solid line) and (b) ZEUS parameterization (dashed line).
}
\end{figure}

\noindent
In case $(B)>(A)$ an algorithm is obviously too inaccurate and corrupts 
result more than uncertainties with
the muon cross-sections and medium composition. $(B)\ll (A)$ 
is better
but leads to unreasonable wasting of human and computer efforts, the
tool for the muon transport is too ``thin'' comparing to insurmountable 
uncertainties of group $(A)$. Thus, the case $(B)<(A)$ seems to be an
optimum equilibrium. 

This report is aimed to get some quantitative parameters of muon transport
algorithm which condition $(B)<(A)$ leads to.      

\section{How precisely do we know the muon energy losses?}

The accuracy of existing formulae for muon cross-sections was analyzed in 
number of works \cite{kp1},\cite{rhode1},\cite{kp2}. 
The conclusion is that it is not higher than $\approx$1\% 
for $E_{\mu} <$ 1 TeV and becomes the worse the higher muon energy is. We do 
not give here the complete analysis but consider only two examples that 
concern the muon photonuclear interaction which contributes the largest 
uncertainty to the total muon energy losses comparing with other kinds of muon
interactions that represent purely electromagnetic (and hence, better known) 
processes: ionization, bremsstrahlung, and direct $e^{+}e^{-}$-pair production.

For last twenty years one has been using the formula for photonuclear
muon cross-section that was developed in the frame of generalized vector 
dominance model (GVDM) \cite{phnubb}. In such an 
approach the differential cross-section (and, consequently, the muon energy 
losses due to photonuclear interactions) is proportional to the total 
cross-section for absorption of a real photon by a nucleon, 
$\sigma^{\gamma p}_{tot}$. There are several parameterizations for 
$\sigma^{\gamma p}_{tot}$ that were obtained by trying to get the best fit 
to experimental data \cite{phnubb},\cite{allm},\cite{dl},\cite{ZEUS},
\cite{butmikh}. Two of them are shown 
in fig.1 along with the last results from
the ZEUS experiment \cite{ZEUS}.
$\sigma^{\gamma p}_{tot}$ is presented as a function of $W^{2}$, where $W$ is 
the center-of-mass energy of the $\gamma p$ system. Parameterizations 
\cite{phnubb} (BB) and \cite{ZEUS} (ZEUS) 
both fit more or less the experimental data but even within 
``experimentally known'' range $W^2 \leq$ 10$^5$ GeV$^2$
the difference between 
two parameterizations reaches 4\%.
At $W^2$ = 10$^8$ GeV$^2$ the BB curve is higher than the ZEUS one by 15\%.
The same data are presented in fig.2 in terms of muon energy
losses due to photonuclear interaction in pure water. Difference is kept
within several percents below $E_{\mu} \approx$ 100 TeV and then grows
up to $\approx$ 10\% 
at $E_{\mu}$ = 1 EeV.  
Parameterizations for $\sigma^{\gamma p}_{tot}$
published in \cite{allm},\cite{dl} are lower comparing to ZEUS
parameterization since they are based on old results which differ remarkably 
from the latest ZEUS data. On the other hand, the recent parameterization 
\cite{butmikh} is higher comparing to BB.

\begin{figure}[h]
\vspace*{-2.0mm} 
\includegraphics[width=8.0cm]
{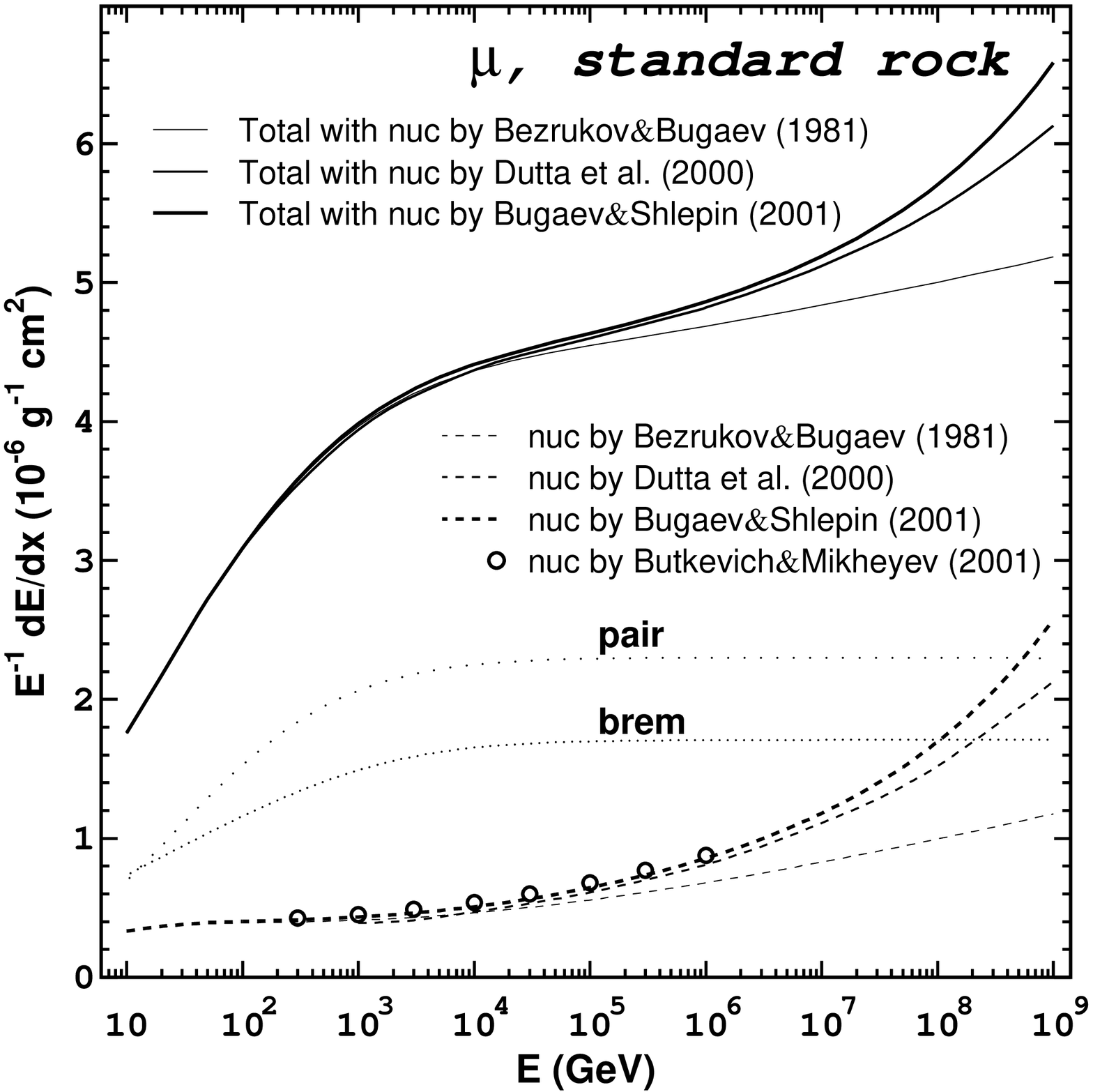} 
\label{fig:qcd}
\caption{
Muon energy losses in standard rock
due to photonuclear interaction with- 
\cite{dutta},\cite{butmikh},\cite{bush}
and without \cite{phnubb}
accounting for QCD part (dashed lines and open circles). Curves for 
bremsstrahlung \cite{brembb1},\cite{brembb2},\cite{bremkok},
direct $e^{+}e^{-}$-pair production 
\cite{pkok1},\cite{pkok2},\cite{pkok4},\cite{pkok3} (dotted lines) are shown, 
as well. Solid lines stand for total energy loss  (brem + pair + nuc, 
ionization not included) with-
\cite{dutta},\cite{bush}
and without \cite{phnubb}
accounting for QCD part in photonuclear interaction.
}
\end{figure}

It became clear in last few years that at very high lepton energies the
essential part of photonuclear cross-section is due to nonGVDM contribution. 
It appears, that this additional part is well described by QCD perturbation 
theory. Recent new calculations \cite{dutta}, \cite{butmikh}, \cite{bush} show
an essential increase of the muon energy losses due to photonuclear 
interactions. In \cite{dutta} the ALLM formulae \cite{allm} (that is based on 
Regge approach and on H1 and ZEUS data) is used for parameterization of 
nucleon structure function $F_{2}$. In \cite{butmikh} the CKMT Regge model 
\cite{capella} is applied for a description of $F_{2}$ at low and intermediate
$Q^{2}$ (squared four-momentum transfer) and the fit of parton distribution 
functions given by MRS group \cite{martin} at high $Q^{2}$. Authors of 
\cite{bush} use for calculation of QCD perturbative part the color dipole 
model in version of \cite{forsh}, with parameters founded from latest DESY 
data. 

\begin{table}
\begin{center}
\begin{tabular}{|c|c|c|c|}\hline
$E_{\mu}$&{\scriptsize Dutta et al.}&{\scriptsize Bugaev\&Shlepin}&{\scriptsize Butkevich\&Mikheev}\\
\hline
100 TeV   &   + 10 \%  &   + 16 \% &   + 22 \% \\
1 PeV     &   + 19 \%  &   + 26 \% &   + 29 \% \\
10 PeV    &   + 34 \%  &   + 42 \% &   - \\
100 PeV   &   + 53 \%  &   + 71 \% &   - \\
1 EeV     &   + 81 \%  &   + 119 \% &   - \\
\hline
\end{tabular}
\caption{Differences in terms of the muon energy losses in standard rock
due to photonuclear 
interaction (in percents) between three works that account for QCD part
\cite{dutta},\cite{butmikh},\cite{bush}
and Bezrukov-Bugaev formula based on GVDM \cite{phnubb}.
}
\label{tab:t1}
\end{center}
\end{table}
\begin{table}
\begin{center}
\begin{tabular}{|c|c|c|}\hline
$E_{\mu}$ & Dutta et al. & Bugaev \& Shlepin \\
\hline
100 TeV   &   + 1 \%  &   + 2 \% \\
1 PeV     &   + 3 \%  &   + 4 \% \\
10 PeV    &   + 6 \%  &   + 7 \% \\
100 PeV   &   + 11 \%  &   + 14 \% \\
1 EeV     &   + 18 \%  &   + 27 \% \\
\hline
\end{tabular}
\caption{Differences in terms of total muon energy losses in standard rock
(in percents) 
between two works that account for QCD part
\cite{dutta},\cite{bush}
and Bezrukov-Bugaev formula based on GVDM \cite{phnubb}.
}
\label{tab:t2}
\end{center}
\end{table}

The results of these three works are presented in fig.3 and tables 1,2 in
terms of the muon energy losses in standard rock.
\cite{butmikh} and \cite{bush} are close to each other while the muon 
energy losses 
obtained in \cite{dutta} are essentially lower. We believe that
this is due to the fact that old data \cite{allm}  were used
in \cite{dutta} to fit the parameters of model. Anyway, the total
muon energy losses become higher when accounting for perturbative QCD
part in photonuclear interactions by 1$\div$2\% 
at $E_{\mu}$ = 100 TeV  and by 20$\div$30\% 
at $E_{\mu}$ = 1 EeV~\footnote{Accounting for nonGVDM corrections becomes even
more important for $\tau$-leptons as photonuclear interaction dominates in 
$\tau$'s energy losses at $E_{\tau} >$ 100 TeV}. Fig.4 shows changes in muon
survival probabilities in sea water and standard rock which incorporation of 
nonGVDM QCD-corrections results in~\footnote{Data for this figure, as well
as for fig.5--10, were obtained with the MUM code for $\mu$ and $\tau$-lepton
transport \cite{mum1} that is available by request to authors.}. It is seen 
that for extremely high energies ($E_{\mu} >$ 1 PeV)
effect is very significant.

\begin{figure}[h]
\vspace*{-2.0mm} 
\includegraphics[width=8.0cm]
{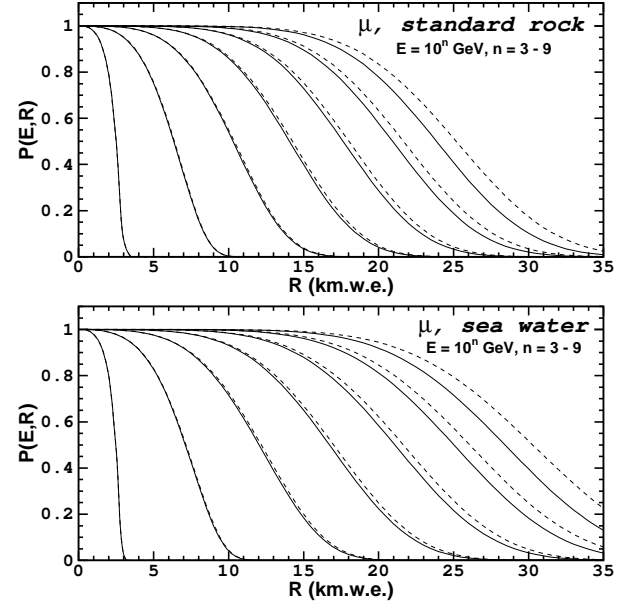} 
\label{fig:tau2}
\caption{
Muon survival probabilities vs slant depth in standard rock (upper panel) 
and sea water (lower panel) when simulating the muon propagation
 with photonuclear interaction 
treated by GVDM approach \cite{phnubb} (dashed lines) and when using also 
correction for QCD part as obtained by \cite{bush} (solid lines). Six pair 
of curves on each panel correspond to six initial muon energies from 1 TeV
up to 1 EeV (from left to the right).
}
\end{figure}

Energy losses are proportional, in particular, to medium density.
The density of water (ice) is known or, in any case, can be measured
precisely. But since the main technique to select the neutrino induced
events is looking for up-going muons from the lower hemisphere, one should
know also the density and composition of sea bed rock. When this
is unknown one has to use a standard rock model
($\rho$ = 2.65 g/cm$^3$, $A$ = 22, and $Z$ = 11). At the same time the
measured rock densities at Gran Sasso and Frejus 
underground laboratories are equal, for
instance, to $\rho_{Gran Sasso}$ = 2.70 g/cm$^3$ and 
$\rho_{Frejus}$ = 2.74 g/cm$^3$,
respectively \cite{macrorock},\cite{frejusrock} which is 2$\div$3\% 
higher. Investigation of the Baikal lake bed (NT-200 neutrino telescope)
showed a complex structure
with several layers of different composition and density that varies with
increase the depth from $\rho_{Baikal}$ = 1.70 g/cm$^3$ to 
$\rho_{Baikal}$ = 2.90 g/cm$^3$ ($\pm$25\% 
variations) \cite{baikalrock}. Thus, error that is incorporated to muon
transport calculations by uncertainties with the rock density/composition
lies, at least, at a few percents level.

All this have led us to a conclusion that we never know the muon
energy losses better than with 1\%-accuracy 
(actually, worse) and, consequently it makes
no sense to waste  efforts trying to reproduce energy losses with accuracy 
better than 10$^{-2}$ with an algorithm for the muon transport. So, there is
a reasonable quantitative criterium for a muon propagation algorithm goodness:
algorithm whose accuracy is better than 1\% 
can be accepted as a good one, otherwise it is bad.  
  
\section{Choice of $v_{cut}$}

Among all simplifications that may be done when developing an algorithm for 
the muon propagation (and that affect the algorithm accuracy) one is 
obligatory. As number of muon interactions per unit of path is practically 
infinite it is impossible to simulate all acts and one has to set some 
threshold for relative energy transfer $v_{cut}$ 
($v$ = $\Delta E_{\mu}/E_{\mu}$, where $\Delta E_{\mu}$ is energy which is 
passed by muon of energy $E_{\mu}$ to either real or virtual photon in a 
single interaction) above which muon energy losses are treated by the direct 
simulation of $\Delta E$ for 
each interaction (``stochastic'' part)
and below which energy losses are treated by 
means of stopping-power formula (``soft'' part) 
that can be obtained by integration of 
differential cross-sections for the muon interactions:
\begin{eqnarray}
\frac{dE_{\mu}}{dx}(E_{\mu}) = \frac{N_{A}}{A}E_{\mu}
\int\limits_{0}^{v_{cut}}\frac{d\sigma(E_{\mu},v)}{dv}v\,dv
\label{cl1}
\end{eqnarray} 
\begin{figure}[h]
\vspace*{1.0mm} 
\hspace{-2.5mm}
\includegraphics[width=8.7cm]
{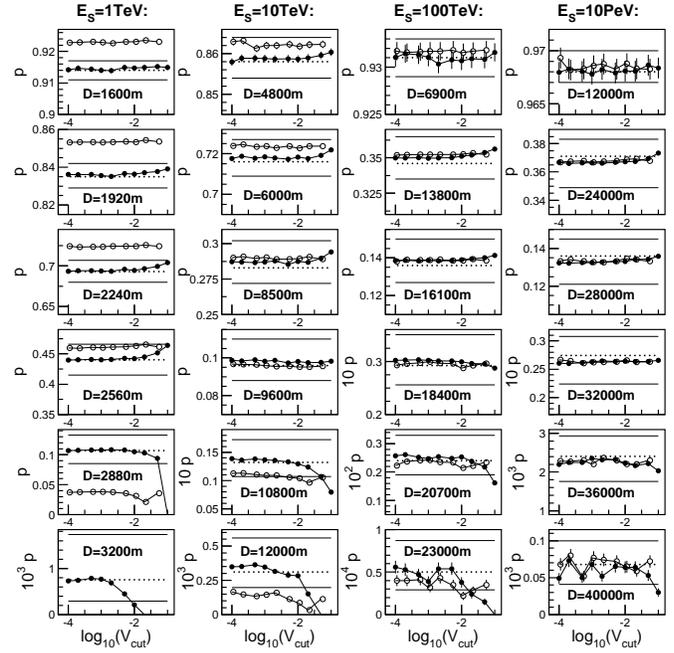} 
\label{fig:fig1}
\caption{
Survival probabilities $p$ vs $v_{cut}$ obtained as a result of MC simulation 
for monoenergetic muon beams with initial energies $E_{s}$ = 1 TeV (1st column
of plots), 10 TeV (2nd column), 100 TeV (3rd column) and 10 PeV (4th column). 
Each column contains six plots which correspond to six depths $D$. See text 
for more details.
}
\end{figure}
\noindent
($A$ is a mass number of the target nucleus, $N_{A}$ is the Avogadro number).
Number of interactions to be simulated per unit of muon path depends on 
$v_{cut}$, growing roughly as $N_{int} \propto v_{cut}^{-1}$ along with 
computation time. So, it would be desirable to choose $v_{cut}$ as large
as possible. But, on the other hand, increase of $v_{cut}$ affects the
simulation accuracy since some part of statistical fluctuations in energy 
losses goes out of simulation and is not accounted for. Thus, the question is 
{\it how large value of $v_{cut}$ may be chosen to keep result within 
1\%-accuracy}? 
This problem was discussed in literature \cite{N94}, \cite{lipari}, 
\cite{music1}, \cite{lagutin1} but, in our opinion, more careful analysis is 
lacking.

Let's firstly consider propagation of monoenergetic muon beams. Fig.5 shows 
results on propagation (survival probabilities $p$ vs $v_{cut}$) for four 
beams with initial energies 1 TeV, 10 TeV, 100 TeV and 10 PeV through slant 
depths 3.2 km, 12 km, 23 km and 40 km, respectively, in pure water.
Closed circles represent results obtained 
with knock-on electrons (that appear as a 
result of ionization with large energy transferred to an atomic electron) 
included in direct simulation, open circles stand for completely 
continuous ionization that is treated simply by Bethe-Bloch 
formula. Solid lines on each panel correspond to $v_{cut}$ = 10$^{-4}$ and all
energy losses multiplied by factors 0.99 (upper line) and 1.01 (lower line), 
thus these lines limit ``$\pm$1\%-energy-losses-error-band''. 
Horizontal dotted lines correspond to $v_{cut}$ = 10$^{-4}$ and cross section 
for absorption of a real photon by a nucleon parametrized according to 
\cite{ZEUS} instead of parameterization \cite{phnubb} that was mainly applied 
at presented simulations.

One can conclude as follows:
\begin{itemize}
\item[(a)] In most cases with except for lower row and the left column (that 
corresponds to low survival probabilities and low muon initial energies, 
respectively) survival probabilities lie mainly within $\pm$1\%-band.
\item[(b)] The difference between survival probabilities for two models of
ionization - partially fluctuating (above $v_{cut}$) and completely ``soft'' -
is the less appreciable the larger muon energy is. 
\item[(c)] Parameterizations \cite{phnubb} and \cite{ZEUS} do not differ 
noticeable from each other in terms of survival probabilities.
\item[(d)] For $v_{cut} \le$ 0.02$\div$0.05 there is almost no dependence of
survival probability on $v_{cut}$ with except for very last part of muon path 
where survival probability becomes small. 
\end{itemize}    

Now let's go to the more realistic case with atmospheric muons. Fig.6 
represents intensities of vertical atmospheric muon flux at 8 depths $D$ 
of pure water as functions of $v_{cut}$ as obtained by simulation with 
muons sampled according to sea level spectrum \cite{bks1}:
{\footnotesize
\begin{equation}
\frac{dN}{dE}=\frac{0.175\:E^{-2.72}}{cm^{2}\:s\:sr\:GeV}
\left(
{1 \over\displaystyle 1 +
{E \over 103\:GeV}}
+
{0.037 \over\displaystyle 1 +
{E \over 810\:GeV}}
\right)
\label{bknsspec}
\end{equation}
}
\noindent
Meaning of closed and open circles, as well as solid and dotted lines is the 
same as for fig.5. Dashed lines on panels for $D \le$ 5 km correspond to 
intensities which were computed for all energy losses treated as completely 
``continuous'' (no fluctuations included). Dash-dotted lines show intensities 
of vertical muon flux simulated with muons sampled according to the Gaisser 
sea level spectrum \cite{gaisser}: 
{\footnotesize
\begin{equation}
\frac{dN}{dE}=\frac{0.14\,E^{-2.7}}{cm^{2}s\,sr\,GeV}
\left(
{1 \over\displaystyle 1 +
{E \over 104.6\,GeV}}
+
{0.054 \over\displaystyle 1 +
{E \over 772.7\,GeV}}
\right)
\label{gaisspec}
\end{equation}
}
\begin{figure}[h]
\vspace*{1.0mm} 
\hspace{-2.5mm}
\includegraphics[width=8.7cm]
{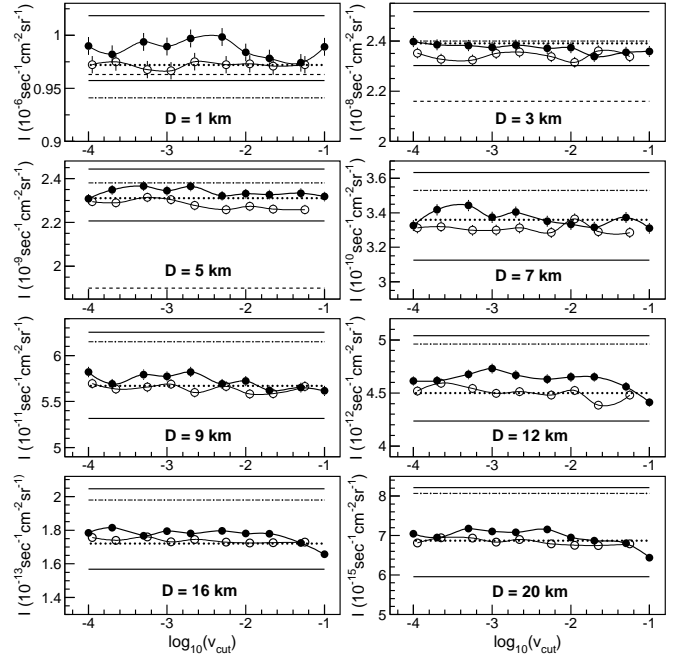} 
\label{fig:fig2}
\caption{
Intensities of vertical atmospheric muon flux at different depths of pure
water vs $v_{cut}$ as obtained by MC simulation. See text for more details. 
}
\end{figure}
General conclusions are qualitatively the same as for mono-energetic muon 
beams but quantitatively the influence of $v_{cut}$ and model of ionization 
is much weaker. One can conclude the following:   
\begin{itemize}
\item[(a)] Computed muon flux is strongly affected by accounting for 
fluctuations in energy losses: muon flux intensity computed by means of 
stopping-power formula is less comparing with MC simulated flux by 
$\approx$10 \%  
at 3 km and by $\approx$ 20\%
at 5 km. At the depth of 20 km muon flux computed with ignorance of 
fluctuations is only 10 \% 
of simulated flux.
\item[(b)] 1\%-uncertainty 
in muon cross sections plays the principal role for resulting error
(all simulated data lie within $\pm$1\%-band with no exceptions).
This error has a tendency to grow with depth 
from $\pm$2.5\% 
at 1 km to $\sim \pm$15\% 
at 20 km. 
\item[(c)] Difference between muon spectra \cite{bks1} and \cite{gaisser} 
leads to uncertainty from \mbox{-4\%}
at 1 km to +16\% 
at 20 km.
\item[(d)] Error which appears due to simplified, entirely ``continuous''
ionization lies, commonly, at the level of 2$\div$3\%.
\item[(e)] Dependence of simulated muon flux intensity upon $v_{cut}$ is the
most weak one comparing with other error sources. 
\end{itemize}

Thus, we can make the following conclusions: to reproduce muon energy losses
with an accuracy better than 10$^{-2}$ it is quite enough to account only for
fluctuations in energy losses with fraction of energy lost being as large as 
$v > v_{cut} \approx$ 0.05$\div$0.1 (commonly, values 
$v_{cut}$ = 10$^{-3}\div$10$^{-2}$  are incorporated in majority of 
algorithms for the muon propagation by now \cite{lipari}, \cite{music1}, 
\cite{mmc1}).
At least, in case with atmospheric muons only radiative 
losses may be simulated while ionization may be treated as an entirely 
continuous process without any lost of accuracy. In spite of remarkable 
dependency of simulated survival probabilities upon accounting or not 
accounting for fluctuations in ionization (see fig.5) it seems to be correctly
to treat ionization by completely continuous way 
also for neutrino-produced muons, as (a) differences in survival 
probabilities appear only for relatively low energies; (b) they are of few 
percents only; (c) they are of both signs and so, are expected to be partially 
self-compensated; (d) the uncertainties with $\nu\,N$ cross-sections are still
much larger, in any case \cite{bartol},\cite{gai1},\cite{lip1},\cite{naumov}. 
With such parameters 
there are only several muon interactions to be simulated per 1 km w.e.     

\begin{figure}[h]
\vspace*{-2.0mm} 
\hspace{-2.5mm}
\includegraphics[width=8.7cm]
{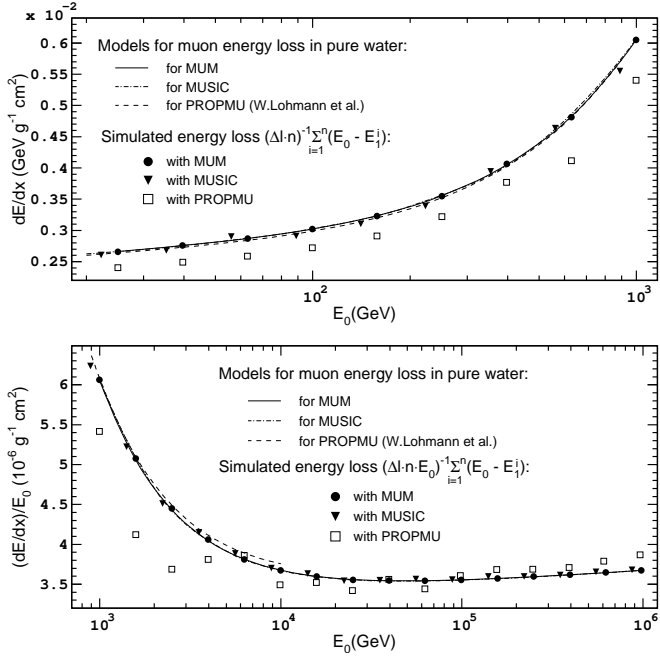}
\label{fig:water}
\caption{Comparison 
between input models for the muon energy losses (lines) and 
simulated energy losses (markers) for muon transport codes 
PROPMU {\it [version 2.01, 18/03/93]}, 
MUSIC {\it [version for pure water with bremsstrahlung cross-sections by 
Kehlner-Kokoulin-Petrukhin, 04/1999]}, 
and MUM {\it [version 1.4, 12/2001]} in pure water.
The energy range 25 GeV$\div$1 PeV is divided into two parts to keep linear
scale: the upper plot stands for energy losses 
vs $E_{\mu}$ = 25 GeV$\div$1 TeV, 
the lower one does for energy losses divided by energy 
vs $E_{\mu}$ = 1 TeV$\div$1 PeV. The model for PROPMU energy losses
is taken from \cite{lohmann} as referenced in \cite{lipari}.
Curves for MUM's and MUSIC's energy loss models are very close to each
other and therefore are hardly distinguished.
}
\end{figure}
\begin{figure}[h]
\vspace*{-2.0mm} 
\hspace{-2.5mm}
\includegraphics[width=8.7cm]
{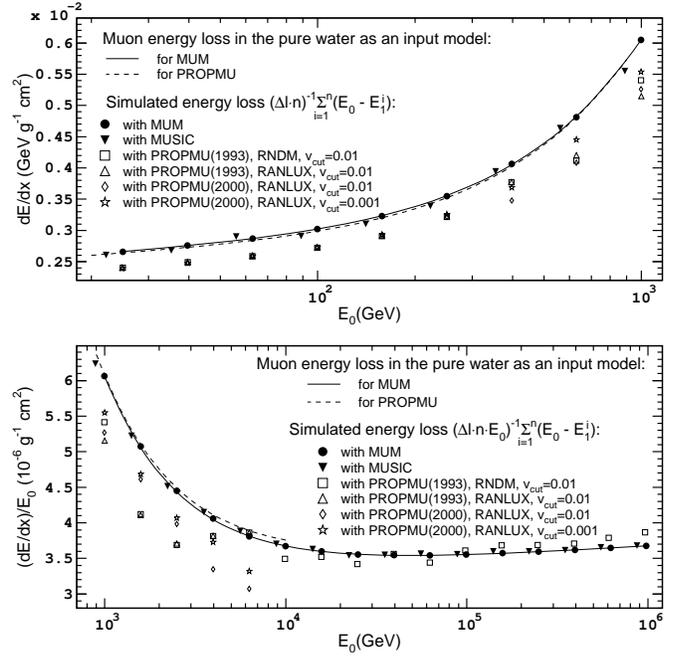}
\label{fig:water_add}
\caption{
The same as in fig.7 with pure water, MUM 
{\it [version 1.4, 12/2001]} and 2 versions of PROPMU 
{\it [versions 2.01, 18/03/93 and 2.1, 01/2000]} with different values
of $v_{cut}$ and different random generators. 
}
\end{figure}
%

We would like to emphasize that uncertainties with atmospheric muon flux 
intensity that are caused by only $\pm$1\% 
variations in energy losses are much higher than 1\% (see item (b) above).
This is due to sharp power-law  shape of surface atmospheric muon spectrum
and due to the fact that source of atmospheric muons is far away from an 
underwater/ice detector and their flux may only decrease when passing from 
the sea level down to detector depth. The source of muons that are produced 
in $\nu N$ interactions is uniformly distributed over water/ice and/or rock 
both out- and inside the array. So, uncertainties with energy losses should
affect the results for neutrino-generated muons weaker. Indeed, intensity of
the muon flux which accompanies the neutrino flux in a medium is proportional 
to the muon range that, in turn, is inversely proportional to energy losses 
and, consequently, an error in energy losses leads to an equal error
with an opposite sign in 
simulated flux of muons which are born in $\nu\,N$-interactions.

\section{``Inner accuracy'' of muon transportation algorithm}

Any muon MC propagation algorithm consists of a set of procedures on 
numerical solution of equations, interpolation and integration. All these 
procedures are of finite accuracy and, consequently, the incoming model for 
muon interactions is somewhat corrupted by them. Having a set of formulae
for the muon cross-sections for bremsstrahlung, $e^{+}e^{-}$-pair production,
photonuclear process, and knock-on electron production it is easy to obtain the
formula for the total averaged 
muon energy losses by integration of differential 
cross-sections over all range of kinematically allowed $v$:
\begin{eqnarray}
\frac{dE_{\mu}}{dx}(E_{\mu}) = \frac{N_{A}}{A}E_{\mu}
\int\limits_{v_{min}}^{v_{max}}\frac{d\sigma(E_{\mu},v)}{dv}v\,dv
\label{losses}
\end{eqnarray} 

\noindent
But energy losses as they are simulated by an algorithm will differ
from (4) due to ``corrupting'' 
application of numerical procedures. Two algorithms
with the same formulae for cross-sections at the input will reproduce the
muon energy losses differently. The difference between simulated and 
calculated energy losses collects all errors that are contributed by each step 
of algorithm and thus, is a good quantitative criterium for its 
{\it inner accuracy}, whose contribution to the resulting error must not 
exceed 1\%, 
as it was shown in the Sec. 2.

\begin{figure}[h]
\vspace*{-7.0mm} 
\hspace{-2.5mm}
\includegraphics[width=8.7cm,height=10.1cm]
{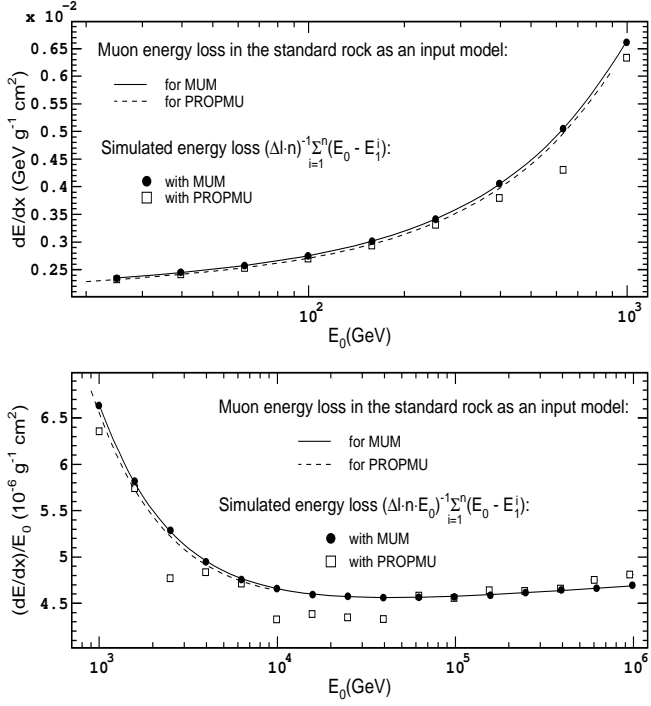} 
\label{fig:strock}
\caption{
The same as in fig.7 and fig.8 with standard rock 
($\rho$ = 2.65 g/cm$^{3}$, $A$ = 22, and $Z$ = 11), MUM 
{\it [version 1.4, 12/2001]} and PROPMU 
{\it [version 2.1, 01/2000]}.
}
\end{figure}
\begin{figure}[h]
\vspace*{-2.0mm} 
\hspace{-2.5mm}
\includegraphics[width=8.7cm]
{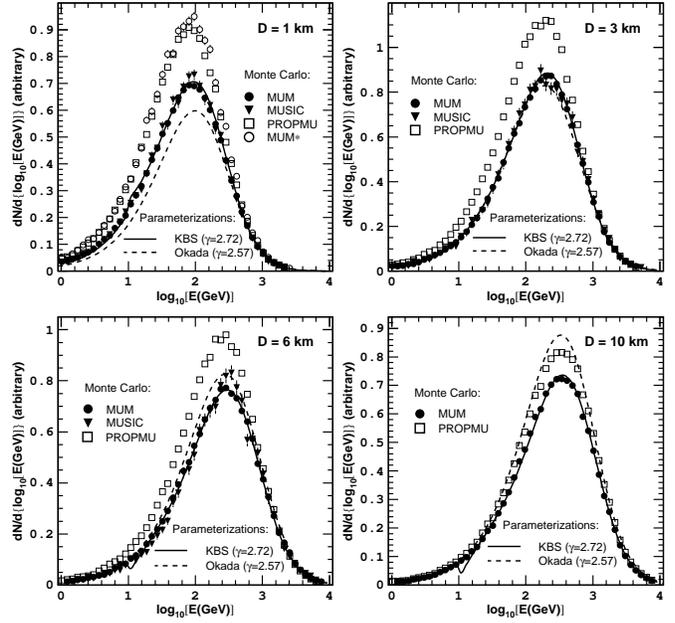} 
\label{fig:dif1}
\caption{
Vertical muon depth spectra in pure water at depths $D =$~1~km, 3~km, 6~km
and 10~km, as simulated with PROPMU, MUSIC and MUM for atmospheric muons
sampled according to sea-level spectrum \cite{bks1}.
For a comparison, parameterizations \cite{bks1} (KBS) and
\cite{okada} (Okada) are given, as well. On panel for $D =$~1~km
the spectrum which resulted from simulation with MUM with all cross-sections
multiplied by 0.9 is also presented (MUM$\ast$, open circles).
}
\end{figure}

The trivial test may be suggested to check inner accuracy of any muon transport
algorithm: let's simulate propagation of $n$ muons (we used $n$ = 10$^{6}$ for
test reported below) with energy $E_0$
over a short distance $\Delta l$ ``measuring'' the final energy $E^{i}_{1}$ 
for each $i-$th muon at the end. Then, simulated energy loss may be computed
as
\begin{eqnarray}
\frac{dE_{\mu}}{dx}(E_{\mu}) = \frac{1}{n\,\Delta l}\;\sum^{n}_{i=1}\;(E_{0}- E^{i}_1)
\end{eqnarray} 

\noindent
Comparing (5) with (4) it is possible to determine a goodness of the given
algorithm applying, for example, criterium 
$abs((5) - (4)) <$ 1\%.

Fig.7, fig.8, and fig.9 show results of such a comparison for three
muon transport codes: PROPMU \cite{lipari}, 
MUSIC \cite{music1},
and MUM \cite{mum1}.
Results for pure water are presented in fig.7. 
In case with MUM the difference $abs((5) - (4))$ never exceeds 1\%, 
for MUSIC it reaches 2$\div$3\%. 
Inaccuracy of the PROPMU code is much worse and in a range $E_{\mu} <$ 10 TeV 
is up to 25\%. 
%
%
We try 2 versions of PROPMU with different values of $v_{cut}$ and different
random generators (fig.8). In some cases inner accuracy becomes better but, 
in any case it stays worse than 10\% 
that is unacceptable according to conclusion of Sec.2. Fig.9 reports the 
results of accuracy test for MUM and PROPMU with standard rock. 
Qualitatively, these results are the same. 

In fig.10 differential spectra for atmospheric muons at different depths 
in pure water are presented as simulated with PROPMU, MUSIC and MUM. Muons at 
the sea level were sampled according to the same spectrum 
\cite{bks1}. Also parameterizations \cite{bks1} and \cite{okada} are shown. 
Simulations with MUSIC
and MUM give practically the same results
because both codes reproduce almost the same energy losses (see fig.7).
MUSIC's and MUM's spectra coincide with parameterization \cite{bks1} that is
based on the same sea level muon spectrum as was used for simulation presented
on the plot and on muon propagation with MUM. Parameterization \cite{okada} is
lower than KBS, MUM and MUSIC (up to 18\% 
in terms of integral muon flux at $D =$~1~km) at relatively shallow depths and
becomes higher starting with $D \sim$~5~km. The reason is that it is based on 
very hard sea-level muon spectrum with index $\gamma =$~2.57 \cite{miyake}, 
while 
KBS parameterization adopts a spectrum with $\gamma =$~2.72. As a result,
there are less muons with 
relatively low energies (which contribute the main bulk to the muon flux at 
low depths) and more muons of high energies which contribute to the muon flux 
at larger depths. The reduced muon energy losses in PROPMU result in the depth
muon spectra which (a) are significantly higher (31\%, 30\%, 27\% and 17\% 
in terms of integral muon flux at the depths $D =$~1~km, 3~km, 6~km  and
10~km, correspondingly) and (b) are expanded to the low energies. 
The spectrum for $D =$~1~km was simulated
also with ``corrupted'' version of MUM: all muon cross-sections along with 
Bethe-Bloch ionization formula were multiplied by a factor 0.9, thus energy 
losses were reduced by 10\% 
and became close to ones that are reproduced by PROPMU~\footnote{The majority 
of muons which compose the spectrum at 1~km  depth have the energy in a range 
$E <$~1~TeV at the sea level. For this range PROPMU reduces the simulated 
energy losses by $\sim$~10\% 
as can be seen in upper panel of fig.7.}. The result is given on upper left 
panel of fig.10 by open circles (MUM$\ast$). A good agreement between PROPMU
and MUM$\ast$ represents a good quantitative cross-check for results on 
PROPMU inner accuracy. 

\section{Conclusions}

The optimum value for inner accuracy of a MC code for the muon transport
is close to 1\%. 
In any case we never know the muon energy losses better. 

To keep such inner
accuracy it is quite enough to simulate the muon interactions with a
fractions of energy lost that are as large as $v > v_{cut} =$ 0.05$\div$0.1,
treating the rest of energy losses (soft part) by means of stopping-power
formula. 

The simple test may be applied to measure the accuracy of an
algorithm for the muon transport by comparison of reproduced energy losses
and incoming model for muon cross-sections. 

Such a test has been 
applied to three muon transportation codes. According to ``1\% 
goodness criterium'' codes MUM and MUSIC can be accepted as accurate enough.
Accuracy of the PROPMU algorithm is insufficient (down to 25\%) 
and depends on medium, version of code and simulation parameters.
It leads, for example, to $\approx$30\% 
overestimation of atmospheric muon fluxes at depths 1$\div$5 km w.e.

{\backsect} 

\end{document}